# Optical and Luminescence Properties of Zinc Oxide

P. A. Rodnyi and I. V. Khodyuk
*St. Petersburg State Technical University, St. Petersburg, 195251 Russia*

**Abstract**—We generalize and systematize basic experimental data on optical and luminescence properties of ZnO single crystals, thin films, powders, ceramics, and nanocrystals. We consider and study mechanisms by which two main emission bands occur, a short-wavelength band near the fundamental absorption edge and a broad long-wavelength band, the maximum of which usually lies in the green spectral range. We determine a relationship between the two luminescence bands and study in detail the possibility of controlling the characteristics of ZnO by varying the maximum position of the short-wavelength band. We show that the optical and luminescence characteristics of ZnO largely depend on the choice of the corresponding impurity and the parameters of the synthesis and subsequent treatment of the sample. Prospects for using zinc oxide as a scintillator material are discussed. Additionally, we consider experimental results that are of principal interest for practice.

## INTRODUCTION

Zinc oxide occupies a special place among wide bandgap semiconductors (GaN, ZnS), which have been actively studied because of an increased need for solid state light sources and detectors in the blue and UV spectral ranges [1, 2]. On the basis of GaN and alloys thereof, light emitting and laser diodes in the visible spectral range (460 nm) were developed. However, ZnO is considered to be more favorable for creating UV light-emitting diodes and laser diodes, since the binding energy of excitons in it is considerably higher (60 meV) than in GaN (25 meV) [2]. Zinc oxide possesses a high radiation, chemical, and thermal resistance; it is widely used in creation of various instruments—in particular, to form transparent contacts of solar cells. Due to its unique optical, acoustic, and electric properties, zinc oxide finds use in gas sensors, varistors, and generators of surface acoustic waves [1]. ZnO single crystals are also used as substrates for obtaining gallium nitride thin films, since both crystals (ZnO and GaN) belong to the wurtzite structural type and the incommensurability parameter of their lattices along the c axis is 1.8% [3]. Recently, powders, films, and ceramics of zinc oxide have been finding use in the scintillation technique [4].

First and foremost, zinc oxide is known as an efficient phosphor [1, 4]. However, during many decades of research, there no consensus has been achieved on the emission mechanism of the crystal. Moreover, there are a number of models of luminescence that contradict each other (see below). The objective of this work was to examine and systematize basic experimental data on the optical and luminescence properties of ZnO and models that describe them. We also considered possibilities of controlling the characteristics of ZnO and prospects for applications of zinc oxide.

## BASIC CHARACTERISTICS OF ZINC OXIDE

Zinc oxide is a direct gap semiconductor with a considerable fraction of ionic bonding. Many particular features of ZnO are determined by the fact that, among the elements of the sixth group, the ionization energy of oxygen is the highest, which leads to a strongest interaction between the Zn$3d$- and O$2p$-orbitals [2, 5]. This accounts for an anomalously wide band gap of ZnO, $E_g = 3.4$ eV, compared to that of ZnS, $E_g = 3.7$ eV (theoretically, the band gap width should increase in the series of chalcogenides ZnTe →ZnSe→ ZnS → ZnO [5]).

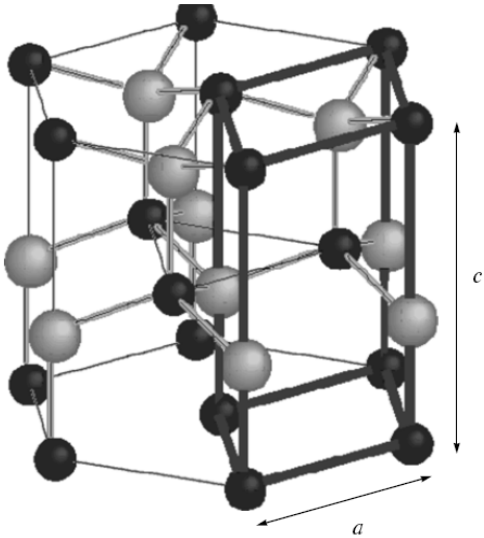
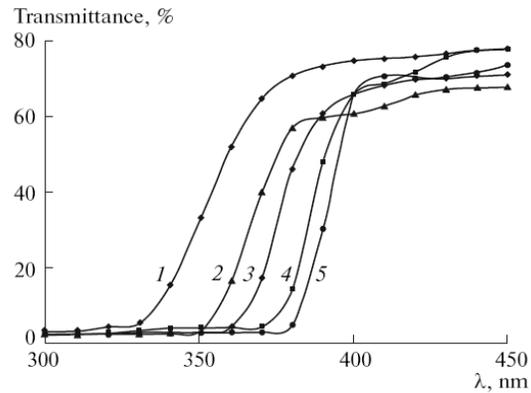

**Fig. 1.** Primitive cell (thick lines) and hexagonal prism of the wurtzite structure of ZnO: (*a*) and (*c*) are the lattice constants (black and white spheres denote Zn and O, respectively).

**Fig. 2.** The shift of the absorption edge of ZnO films in relation to the content Ga (at %): (*1*) 2.5, (*2*) 1.25, (*3*) 0.5, (*4*) 0.125, and (*5*) 0 [6].

Mixing of anionic *p*- and cationic *d*-orbitals in ZnO leads to a low position of the bottom of the conduction band [5].

Under ordinary conditions, ZnO crystallizes in the wurtzite structure with a hexagonal cell (Fig. 1), the space symmetry group is ($P63mc$). Each zinc atom is surrounded by four oxygen atoms, which are located at the corners of a nearly regular tetrahedron. The zinc–oxygen spacing along the *c* axis ($dZn–O = 0.190$ nm) is slightly shorter than that of the other Zn–O bonds ($dZn–O = 0.198$ nm).

Other characteristics of ZnO are as follows: the melting point is 1975°C, the density is 5.67 g/cm3, thelattice constants are $a = 3.25$ and $c = 5.207$ Å, the band gap width is $Eg = 3.437$ eV (1.6 K) and 3.37 eV (300 K), the exciton binding energy is 60 meV, the exciton diameter is ~2 nm; the energy of longitudinal optical phonons is 72 meV, the low-frequency dielectric permittivity is $\varepsilon_0 = 8.75(\parallel)$ and $7.8(\perp)$, and the high frequency dielectric permittivity is $\varepsilon\infty = 3.75(\parallel)$ and $3.70(\perp)$.

At high pressures, a cubic structure is formed (NaCl lattice); cubic ZnO is an indirect gap semiconductor with a band gap width of $Eg = 2.7$ eV [2].

Doping of ZnO films with certain elements (Ga, In, Mg) leads to an increase in the band gap width, as well as to an increase in the activation energy of donor centers and to their stability. An increase in the band gap width of ZnO:Ga compared to ZnO can be judged from the shift of the optical absorption edge of crystals (Fig. 2) [6].

In the majority of cases, zinc oxide is used in the form of thin films and powders or in the form of small size single crystals. Obtaining large single crystals, which are necessary for the scintillation technique, is difficult. The ordinary method of growing crystals from melt does not yield satisfactory results in the case of ZnO because of a high pressure of the vapor [2]. Mainly, large crystals are obtained using the hydrothermal growth method [3]. Hydrothermal growing of crystals is an expensive and prolonged process. Thus, crystals with a size of 30 × 30 × 12 mm were grown for 100 days [3]. In addition, such crystals inevitably contain ionic impurities of K, Li, and other uncontrollable metals. Optical ceramics, which are prepared by the methods of uniaxial hot pressing [7] and hydrostatic pressing [8], serve as an alternative to crystals.

**LUMINESCENCE CHARACTERISTICS OF ZINC OXIDE**

As-grown ZnO crystals obtained by various methods and having a size that varies from tens of nanometers to a few centimeters possess conductivity of the $n$ type, i.e., contain shallow donors, the origin of which remains to be discussed [9]. As a rule, various forms of ZnO, such as single crystals, thin films and threads, nanocrystals, needles, etc., exhibit two luminescence bands—a short-wavelength band, which is located near the absorption edge of the crystal, i.e., the edge luminescence, and a broad long-wavelength band, the maximum of which usually is in the green spectral range (Fig. 3). The edge luminescence, the maximum of which is at 3.35 eV and the decay time is ~0.7 ns, has an excitonic nature [10]. As far as the green luminescence is concerned, despite a vast number of investigations, its nature is yet to be understood. The following luminescence centers were assumed to be responsible for the green luminescence: impurity $Cu^{2+}$ ions [11, 12], zinc vacancies $V_{Zn}$ [13, 14], oxygen vacancies VO [14–17], interstitial zinc ions $Zn_i$ [18], oxygen antisites ZnO [19], and transitions $Zn_i \rightarrow V_{Zn}$ [20]. As a result, the authors of review [20] came to the conclusion that various centers may be involved in the green luminescence simultaneously. This situation obtains due to the imperfection of zinc oxide crystals, the instability of certain point defects, and the variety of their forms.

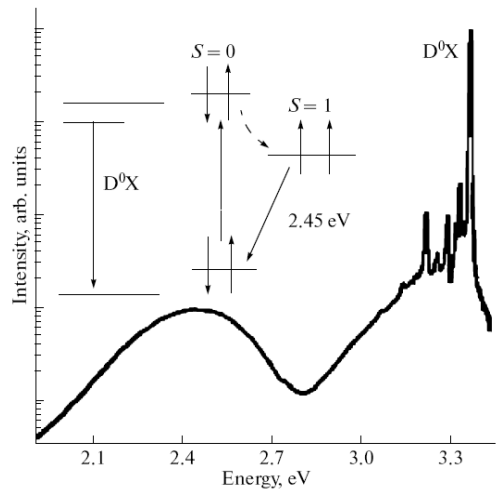

**Fig. 3.** Photoluminescence spectrum of a ZnO crystal and a model of electronic transitions proposed in [16, 17].

Data on the position of levels of point defects in the band gap of the crystal also differ. The point is that, in the wurtzite lattice, interstitial zinc ions $Zn_i$ and oxygen ions $O_i$ may be in an octahedral ($Zn_i$(oct), $O_i$(oct)) and in a tetrahedral ($Zn_i$(tet), $O_i$(tet)) environment. It is known that oxygen interstitials are always more stable in the octahedral environment $O_i$(oct) [13]. We also note that, according to the data of [13], oxygen vacancies VO should predominate in Zn-rich crystals, since their formation energy is lower than that of $Zn_i$ interstitials. In crystals obtained in the oxygen atmosphere, zinc vacancies $V_{Zn}$ should predominate. For a long time, the *n* type of the conductivity of ZnO was attributed to the occurrence of uncontrollable impurities. Recent experiments showed that the interstitial zinc Zn$i$, which is a shallow donor, considerably contributes to the conductivity of the *n* type in ZnO [21].

*Green Luminescence*

Let us consider basic models of the green luminescence and luminescence centers that can be responsible for it, discarding unlikely centers such as interstitial zinc ions Zn$i$ (shallow donor) and oxygen anitisites ZnO, the formation energy of which is high (these anitisites more frequently occur in A3B5 semiconductors).

Certain authors assume that the green luminescence band with a maximum at λ$m$ = 535 nm (2.32 eV) and full width at half maximum Δ$E$1/2 = 330 meV is caused by copper ions, which replace zinc and always occur in ZnO in a small amount [11, 12]. This is confirmed by the similarity of the luminescence bands of undoped ZnO and ZnO:Cu. In addition, equidistant maxima with a step of 72 meV (the energy of longitudinal optical phonons in ZnO) were observed at low temperatures on the short-wavelength wing of the green luminescence band in ZnO and ZnO:Cu.

Subsequent experiments showed that there is no correlation between the concentration of copper ions and intensity of the green luminescence [16, 17]. It was revealed that the green luminescence of ZnO samples annealed in the oxygen atmosphere or in air is related to intrinsic defects, zinc vacancies VZn. The calculations of [13] of the level positions of different defects in ZnO also confirm that the green luminescence arises as a result of recombination of electrons of the conduction band by VZn-centers. It was shown that the zinc vacancy VZn in ZnO is an acceptor, the ground level of which lies 0.8 eV higher than the valence band top, while luminescence centers in *n*-type semiconductors are usually acceptors.

It was shown in [22, 23] that the green luminescence is caused by electronic transitions between shallow donors and deep acceptors (VZn), and the mechanism of the green luminescence in ZnO is similar to that of the yellow band in GaN. At helium temperatures, equidistant maxima at the short-wavelength edge of the green luminescence band have a doublet structure, which is related to the occurrence of two types of shallow donors in ZnO with a depth of location of 30 and 60 meV [22]. Later, this model was refined. The experiment was performed with thin (~300 nm) films of zinc oxide with addition of nitrogen (ZnO:N) [23]. An anomalous temperature dependence of the intensity of green luminescence was observed: the luminescence intensity increased as the temperature of the sample was increased from 15 to ~50 K. An analysis of the obtained data made it possible to assume that, instead of shallow donors of two types, there exist donors of only one type but with two levels, the ground level and the excited one. Therefore, according to [23], the green luminescence occurs as a result of electronic transitions from the ground (D0) and excited (D*) states of the shallow donor to the deep acceptor VZn (on the left in Fig. 4).

The authors of [14] also used the model of the green luminescence with the participation of electronic transitions from the conduction band (or from the donor level) to VZn-levels. The authors of that work proceeded from the fact that the calculated position of the level of the VO-center is 2.7 eV higher than the top of the valence band (therefore, VO cannot be luminescence centers). We note that, at present, based on a number of experimental data for the ground level of the oxygen vacancy, it is agreed that the level of the VO-center is 0.9 eV higher than the top of the valence band [24] and, therefore, VO can be luminescence centers.

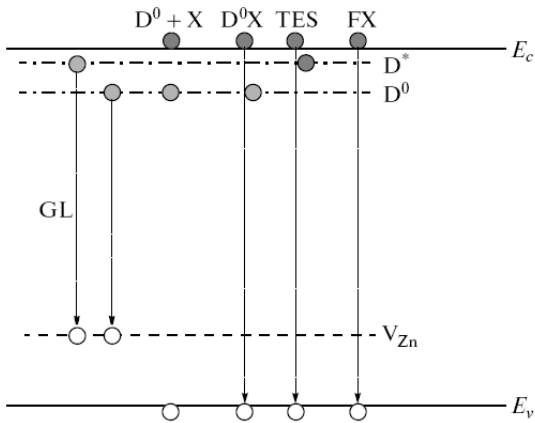

**Fig. 4.** Energy levels of centers responsible for the green luminescence (GL) and the edge luminescence in the band scheme of ZnO (the model of $V_{Zn}$-centers) [23].

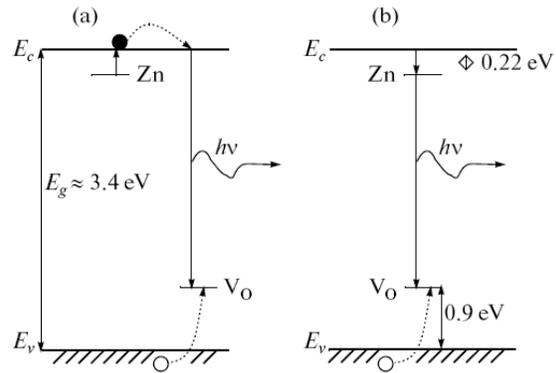

**Fig. 5.** Level positions and electronic transitions responsible for the green luminescence (the model of $V_O$-centers) at (a) high (~300 K) and (b) low (<50 K) temperatures [27].

In [15], thin ZnO films deposited on sapphire substrates were studied.

Based on the data of annealing of samples in nitrogen, oxygen, and air atmosphere, it was concluded that oxygen vacancies are responsible for the intraband luminescence at 510 nm (2.43 eV). The model of oxygen vacancies as green luminescence centers was reliably confirmed by the optically detected magnetic resonance (ODMR) technique [16, 17]. High purity grade single crystals from Eagle-Picher were studied, which contained shallow donors at a concentration of $1 \times 10^{17}$ cm$^{-3}$. It was found that the intensity of the green luminescence ($E_m$ = 2.45 eV, $\Delta E_{1/2}$ = 320 meV) remains high up to 450 K, which is characteristic of deep point defects. At low temperatures, the luminescence line with a maximum at 3.336 eV was also observed (Fig. 3), which is attributed to excitons bound to shallow donors (D0X). By analyzing the ODMR data, the authors of [16, 17] showed that, in the ZnO crystal, energy is transferred from excitons (D0X) to oxygen vacancies (VO-center). Under the action of the incident UV radiation, a neutral oxygen vacancy, which contains two electrons, passes to an excited singlet state and then relaxes to an excited triplet state ($S$ = 1), from which the center luminesces (Fig. 3). It was shown that VO-centers are analogs of $F$-centers, which have been well studied in ionic crystals CaO, SrO, BaO, and MgO.

Experimental detection of centers with $S$ = 1 cannot be doubted; however, the interpretation of the result has been a source of dispute. Some arguments were put forward in theoretical work [25], in which it was shown that the formation energy of oxygen vacancies VO is high and their amount in ZnO cannot be large. The authors of [25] follow the model of $V_{Zn}$-centers of the green luminescence. Performing EPR, photo-EPR, and ODMR experiments with ZnO crystals irradiated by electrons, the authors of [26] came to the conclusion that the signal with $S$ = 1 of the ODMR spectrum should be attributed to more complicated complex centers, which involve VO vacancies, rather than to single VO vacancies.

The study of the temperature dependence of the maximum position of the green luminescence of ZnO nanoneedles showed that the following two processes contribute to the luminescence emission: (i) high-temperature recombination of electrons of the conduction band with holes at the VO-level and/or (ii) low-temperature recombination of electrons of Zn$i$-centers with holes at the VO-level [27] (Fig. 5). At high temperatures (~300 K), thermal transfer of an electron from the Zn$i$-level to the conduction band occurs and then its recombination takes place with a hole at the VO-level (Fig. 5a). As a result, at 300 K, the maximum of the green luminescence band is located at 2.25 eV, with its halfwidth being 600 meV. At low temperatures, the electron from the conduction band is captured at the Zn$i$-level and then recombines with a hole at the VO level (Fig. 5b). As a result, at 50 K, the maximum of the green luminescence band is located at 2.1 eV, with its halfwidth being 400 meV. The authors of [27] consider their experimental results to be consistent with the ODMR data from [17].

In [28], a ZnO single crystal with a high resistivity (~2 kOhm cm) that was obtained by introducing a small amount of lithium ions was investigated. To determine whether VO- or VZn-centers are responsible for the green luminescence, samples were annealed at a high temperature (up to 1050°C) either in oxygen atmosphere or in zinc vapor. It was shown that, in samples with an excess of oxygen, the maximum of the green luminescence band is located at 2.35 ± 0.05 eV, and zinc vacancies VZn are responsible for the luminescence. In samples with an excess of zinc, the maximum of the green luminescence band is located at 2.53 ± 0.05 eV and the luminescence centers are oxygen vacancies VO. Clearly, in O-excess samples, interstitial oxygen centers O$i$ are also formed, while, in Zn excess samples, interstitial zinc centers Zn$i$ are formed with centers of the two types forming shallow levels in the crystal.

ZnO samples doped with lithium exhibit a yellow luminescence band ($Em$ = 2.20 eV, $\Delta E1/2$ = 500 meV) [2, 20]. The lifetime of the band is long, and the band is also observed in thermally stimulated luminescence. Zinc-replacing Li creates an acceptor level in the crystal, which is ~0.8 eV higher than the top of the valence band. The yellow luminescence is attributed to recombination of donors with lithium acceptors [20, 28].

A red luminescence band ($Em$ = 1.75 eV, $\Delta E1/2$ = 500 meV, $\tau \approx 1$ μs) appears in the spectrum of an undoped ZnO crystal upon its annealing in air at 700 K for an hour [20]. The intensity of the red band significantly decreases (more than by an order of magnitude) as the crystal is heated from 15 to 300 K.

Analysis of the data considered above suggests that, in zinc oxide with an excess of zinc (ZnO:Zn), oxygen vacancies VO or (in terms of ionic compounds) $F$-centers are responsible for green luminescence. Since the $F$-center in ZnO contains two electrons, its states are similar to those of an autolocalized exciton in alkali halide crystals; i.e., we are dealing with exciton-like triplet-singlet luminescence [29]. Centers can be excited either via exciton states (D0X) [16, 17], or via donor Zn$i$-centers, or recombinatively [27]. In samples with an excess of oxygen, VZn-centers are responsible for the green luminescence.

*Edge Luminescence*

The deexcitation time of edge luminescence is in the subnanosecond range; therefore, this phenomenon is most important for high-speed devices (lasers, scintillators, phosphors). Edge luminescence involves the participation of free excitons (FXs), excitons bound to acceptors and donors and their two-electron satellites, and donor–acceptor pairs; in addition, phonon replicas are detected. The whole variety of lines belonging to centers described above were described in detail in [10] (exciton lines are also seen in the high-energy range of the luminescence spectrum presented in Fig. 3). Here, we will consider the experimental data on the study of the edge luminescence that are of the most interest for practice.

In [30], the parameters of the edge luminescence in ZnO crystals, nanocrystals with an average particle size of 20 nm, and quantum dots with a particle size of 4 nm, which is approximately twice as large as the exciton diameter (~2 nm) in ZnO, were compared. In photoluminescence spectra of samples, radiation of D0X_excitons with a maximum at 3.25 eV at 300 K predominates. The luminescence intensity increases with decreasing particle size, while the band halfwidth changes as follows: 172, 95, and 85 meV in quantum dots, nanocrystals, and crystal, respectively.

The temperature quenching of luminescence is described by the well-known formula

$$I(T) = \frac{I_0}{1 + Ae^{-E_a/kT}}, \qquad (1)$$

where $I(T)$ is the luminescence intensity at the temperature $T$, $I_0$ is the initial intensity at low temperatures, $E_a$ is the activation energy, and $A$ is a constant. In studies of the temperature quenching of the edge luminescence, the following values of the activation energy were obtained: for the crystal, $E_a = 59.0$ eV [30], and, for thin films, $E_a = 59.7$ eV [31]; both values agree well with the exciton binding energy of 60 meV in ZnO.

The temperature positions of the exciton peak maxima $E_m(T)$ follow changes in the band gap width of the crystal and is determined by the Varshni formula [32]

$$E_m(T) = E_m(0) - \frac{\alpha T^2}{T + \theta_D}, \qquad (2)$$

where $E_m(0)$ is the maximum of the line at $T = 0$ K, $\theta_D$ is the Debye temperature (for ZnO, it was assumed that $\theta_D = 920$ K) and $\alpha$ is a constant. For D0X_excitons in the crystal, it was found that $E_m(0) = 3.360$ eV and $\alpha = 0.67$ meV/K [28].

In thin ZnO films grown on the (111)_surface of CaF2 at 10 K, the photoluminescence line with $E_m = 3.366$ eV predominates and $\Delta E1/2 = 12$ meV, for which excitons that are bound to neutral donors D0X are responsible [33]. Upon heating to room temperature, natural broadening of the line is observed and its maximum shifts to the range ~3.29 eV.

The effect of the concentration of free electrons (in the range from 1013 to 1018 cm–3) on the position of the maximum of the edge luminescence of ZnO crystals was studied in [34]. It was shown that, at 300 K, the photoluminescence maximum is shifted from 3.312 eV at $n_e = 1013$ cm–3 to ~3.27 eV at $n_e = 1018$ cm–3. Based on the data on low values of $n_e$, the band gap width of the crystal was determined to be $E_g = 3.372$ eV. It is assumed that the shift of the FX-exciton maximum toward low energies with increasing $n_e$ is associated with a decrease in the band gap width.

*Interrelation between Green and Edge Luminescence*

That the green luminescence is interrelated with the edge luminescence has been noted by advocates both of the model of zinc vacancies [23] and of the model of oxygen vacancies [16] for green luminescence centers. Basic radiative transitions in ZnO in which VZn serve as luminescence centers are shown in Fig. 4 [23]. As was noted, the green luminescence arises as a result of electronic transitions from the ground (D0) and excited (D*) states of a shallow donor to a deep acceptor VZn. For the edge luminescence, in Fig. 4, we also noted transitions that correspond to the luminescence emission of free excitons (FX), of excitons bound to neutral donors (D0X), and of their two-electron satellites (TES). The electronic states D* and D0X are considered as an intermediate and the ground (initial) states, respectively. At a low temperature, electrons of the ground state do not participate in the green luminescence; however, with increasing temperature, bound excitons decay according to the reaction D0X → D0 + FX, which leads to the participation of neutral donors in radiative transitions D0 → VZn and to an increase in the intensity of the green luminescence. TES can also transfer electrons to an excited donor level (Fig. 4), i.e., contribute to the intensity of the green luminescence.

In ZnO samples with luminescence centers in the form of oxygen vacancies, energy transfer from excitons to VO also occurs [17]. The exciton spectrum of the green luminescence contains a broad band (excited states of VO vacancies) with a maximum at 3.5 eV and a halfwidth of 300 meV and a sharp maximum in the range of formation of bound excitons (~3.35 eV) [17].

## CHANGING THE CHARACTERISTICS OF ZINC OXIDE BY EMBEDDING DONORS AND ACCEPTORS

Since edge luminescence can involve donors and acceptors, it is possible to control the position of the edge luminescence band by choosing corresponding impurities. The energy of photons that arise upon recombination of free electrons with acceptors (A, e) or holes with donors (D, h) is lower than the energy of free excitons. Using the (A, e)_ or (D, h)_luminescence, one can implement a red shift of the edge luminescence band and reduce its overlap with the crystal absorption edge.

It is known that the concentration of free electrons in ZnO can be increased by introducing trivalent impurities Al, In, and Ga [10]. Difficulties arise upon obtaining conductivity of the *p*-type. This is performed by introducing elements of the fifth group (N, P, As, Sb), which replace oxygen, or elements of the first group (Li, Na, K), which should replace zinc. Attention should be paid to the difference between the requirements to donors and acceptors with respect to (i) obtaining the conductivity of the *n*- or *p*-type and (ii) the red shift of the edge luminescence band. In the first case, shallow (<30 meV) donors and acceptors are required. In the case of (D, h)_recombination, the ionization energy of the donor should be in the range 50 meV ≤ $E_i$ ≤ 200 meV. The upper bound ensures efficient electron capture from the conduction band, and the lower bound is required to avoid the ionization of the center prior to its involvement in the emission process. On the contrary, acceptors available in these crystals should have shallow levels (<30 meV), i.e., a high probability of thermal ionization (at 300 K, the energy is $kT \approx 25$ meV). Clearly, in the case of recombination of electrons with acceptors, the position of the acceptor level should satisfy the requirement 50 meV ≤ $E_i$ ≤ 200 meV.

*Acceptors*

Among the elements of the first group, Li is considered the best acceptor impurity, since the bond length $d$Li–O = 2.03 Å is the closest to average the bond length Zn–O 1.93 Å (for other ions, the bond length is longer: $d$Na–O = 2.10 Å, $d$K–O = 2.42 Å) [35]. However, since the ionic radius of Li+ is small, Li ions easily migrate over the crystal and occupy interstitial positions. Upon introduction of elements of the fifth group, difficulties are also encountered and the bond length $d$ of P and As is large; therefore, they form antisites. Nitrogen is the best acceptor (it occupies oxygen sites) with the bond length $d_{N–Zn}$ = 1.88 Å; in addition, the ionization energy of nitrogen is low [35]. Nitrogen is introduced in the form of N2, NO, NO2, and NH3. However, the solubility of nitrogen in ZnO is low; therefore, it is often introduced with the ion-implantation method. Theoretical consideration of the problem showed that the edge luminescence band can be redshifted upon replacement of oxygen by fluorine or upon replacement of zinc with magnesium or beryllium [36].

In ZnO:N films, the concentration of holes of $9 \times 10^{16}$ cm$^{-1}$ has been obtained [37]. A photoluminescence peak at 3.32 eV was observed in samples, which was attributed to the radiation of (A0, X)_centers. The location depth of acceptors was 170–200 meV.

*Donors*

The effect of gallium on the characteristics of the edge luminescence was studied using ZnO:Ga films obtained by the laser molecular epitaxy method [38]. As the concentration of Ga was increased from $8 \times 10^{18}$ to $6 \times 10^{20}$ cm$^{-3}$, the maximum of the edge luminescence band ($E_m$) was shifted from 3.27 to 3.34 eV ($T$ = 293 K). Characteristically, the band gap width increases faster with increasing Ga concentration than $E_m$; as a result, at $n_{Ga} = 6 \times 10^{20}$ cm$^{-3}$, it was found that $E_g$ = 3.7 eV, and the Stokes shift is $S \approx 400$ meV. It is believed that the luminescence radiation arises due to the recombination of donor (gallium) electrons with valence holes; i.e., we are dealing with the (D, h)_luminescence. As $n_{Ga}$ increases from $8 \times 10^{18}$ to $1.5 \times 10^{20}$ cm$^{-3}$, the edge luminescence intensity increases 17_fold. Then, at $n_{Ga} > 1.5 \times 10^{20}$ cm$^{-3}$, a decrease in the intensity is observed, which was attributed to the radiation of donor–acceptor pairs [38]. However, it is known that the radiation of donor–acceptor pairs should obey a hyperbolic law, whereas the experimentally observed decrease was exponential. In [39], it was noted that the edge luminescence is asymmetric and contains phonon replicas.

Upon laser excitation of a ZnO:Ga crystal (0.01%), the luminescence decay time was observed to be very short, 44 ps; in this case, the edge luminescence predominated ($\lambda_m$ = 390 nm), while the intensity of the green luminescence was very low [40]. We also studied the characteristics of the crystal in the case of excitation with α-particles, protons, and X rays (the ZnO:Ga–0.01% crystal with a diameter of 50 mm and a thickness of 50 μm was obtained by the magnetron-injection method) [40].

*Double Donor–Acceptor Doping*

An even greater red shift of the edge luminescence can be achieved if donors and acceptors are simultaneously introduced into ZnO for obtaining the radiation of donor–acceptor pairs. It was shown in [41] that a 1 : 2 Ga : N ratio in the crystal is optimal. In ZnO(Ga, 2N), stable gallium–nitrogen (N–Ga–N) complexes are formed, which occupy certain lattice sites [41]. In this way, it becomes also possible to increase the general content of nitrogen in ZnO. In compounds ZnO(Ga, 2N) [41], ZnO(Al, 2N), and ZnO(In, 2N) [42], a smaller depth of location of acceptor levels compared to that in ZnO(N) has been obtained. The possibility of codoping of zinc oxide with lithium (Li$^+$-acceptors) and fluorine (F$^-$-donors) was also considered [41].

PRACTICE-ORIENTED STUDIES OF ZnO
*Thin Films, Powders, and Single Crystals*

Use as a scintillator is an innovative application for zinc oxide. In this case, the following characteristics of ZnO are important: a short deexcitation time of the excitonic band, a high mechanical and chemical stability, a high radiation resistance, and a rather high density ($\rho$ = 5.61 g/cm3). Microsized powderlike ZnO:Ga possesses a high specific light yield, 15000 photons/MeV, and a deexcitation time of 0.7 ns; as a result, the quality factor (the ratio between the specific light yield and the decay time) of ZnO:Ga is the highest among all available phosphors [4].

A ZnO:Ga scintillator is applied as an efficient detector of α-particles in deuterium–tritium (D–T) neutron generators, which are used in customs inspection of containers [43]. Commonly, ZnO:Ga is used in the form of powderlike [44] or thin-film [45] coatings. The authors of [44] examined the spectral and kinetic characteristics of ZnO:Ga powders of different origins and particle sizes. The maximum of the edge luminescence band is slightly shifted from 386 nm (3.21 eV) at an average particle size of $d$ = 0.9 μm to ~390 nm (3.18 eV) for $d$ = 8.2 μm. The best characteristics (high intensity of the edge luminescence and absence of the green luminescence) belong to the ZnO:Ga powder with $d$ = 8.2 μm that is produced at the Lawrence Berkeley Laboratory. It was shown that donor-bound excitons are responsible for the edge luminescence.

It was attempted to increase the edge luminescence intensity by introducing activators H+, N3–, and S2– into ZnO:Ga powders [46]. In samples annealed in hydrogen atmosphere, a maximal intensity of the edge luminescence was achieved at a content of 0.08% of Ga2O3 in ZnO. An anticorrelation between the edge luminescence intensity and the electron conductivity of samples in relation to the content of Ga and annealing temperature was noted. Thin-film ZnO:Ga scintillators were also studied in the case of proton excitation [47]. Undoped ZnO films also show short times of the edge luminescence ($\tau$ = 550–700 ps at 300 K and $\tau$ = 60–130 ps at 4 K) upon fast-electron excitation [48].

Thin-film ZnO:In samples were studied upon excitation by α-particles [45]. With increasing In content, the intensity of the visible (~500 nm, 2.48 eV) luminescence decreased, and, for the edge luminescence at ~375 nm (3.31 eV), an optimal content of In was 26 ppm. The specific light yield of samples was ~800 photons/MeV, which constitutes ~10% of the light yield for the standard BGO (bismuth germanium oxide) scintillator.

Good scintillation characteristics were obtained for crystals ZnO:Ga,P, ZnO:Ga,N, ZnO:Ga,S and some other direct gap semiconductors [49]. In [50] a light yield at a level of the BGO light yield was obtained for ZnO:Ga, and, for ZnO:Ge, it was at a level of 120% of the BGO yield, with the decay time of the edge luminescence band at 370 nm (3.36 eV) being <0.6 ns.

The authors of [51] studied the temperature dependence of the edge luminescence band at 386 nm (3.21 eV) of a ZnO crystal excited by a UV laser (13.9 nm). In the temperature range 25–300 K, the decay time of the main component of the edge luminescence ($\tau 1$) remained at a level of 0.7–0.9 ns, whereas the decay time of a slow component ($\tau 2$) increased from 1.8 to 3.7 ns.

At present, a search for ultrafast scintillators for successfully developed X-ray free electron lasers (XFELs), which emit powerful X-ray pulses of a femtosecond and even of an attosecond range, has been performed [52]. Zinc oxide is the best-suited material for obtaining femtosecond scintillations. ZnO single crystals were grown by the hydrothermal method, and, to reduce the decay time, iron was introduced (0.61 ppm) [53]. For the excitation, radiation at a wavelength of 51–61 nm was used and detection was performed with a streak camera. The scintillation risetime in ZnO:Fe was 38 ps, while the decay time consisted of two components (2.1 and 70 ps).

Studies aimed at creating a ZnO-based UV laser are progressing, and induced radiation has been obtained for *p–n*-transitions and in thin films [54].

*Ceramics*

As was noted above, as an alternative to single crystals (growing which in the case of ZnO is a very laborious process), optical ceramics serve, which are obtained by the method of uniaxial hot pressing [7]. This method was previously used in development of scintillation ceramics Gd2O2S:Pr,Ce [55] and ZnO:Zn [56]. ZnO-ceramics were obtained using powderlike zinc oxide with an initial particle size in the range 90–600 nm [57–60]. Figure 6 presents X-ray luminescence spectra ZnO and ZnO:Ga ceramics in comparison with the spectrum of the well_known BaF2 scintillator. In undoped ZnO ceramics, the intraband luminescence of VZn-centers predominates, the maximum of which is at 520 nm (2.37 eV). In the visible range, the transmittance of samples of ZnO ceramics with a thickness of 1.0 mm was 40–45% and the integrated light yield was ~540% of that of BaF2, which allows one to consider ZnO ceramics to be a promising scintillation material. Figure 6 also shows that the emission spectrum of the ZnO ceramics is quite consistent with the spectral sensitivity of modern photodetectors, such as silicon photon multipliers (SiPMs) and CCDs. In the luminescence kinetics of the ZnO ceramics (Fig. 7a), one can distinguish fast and slow components with decay constants of 13 ns and 1.6 μs, respectively.

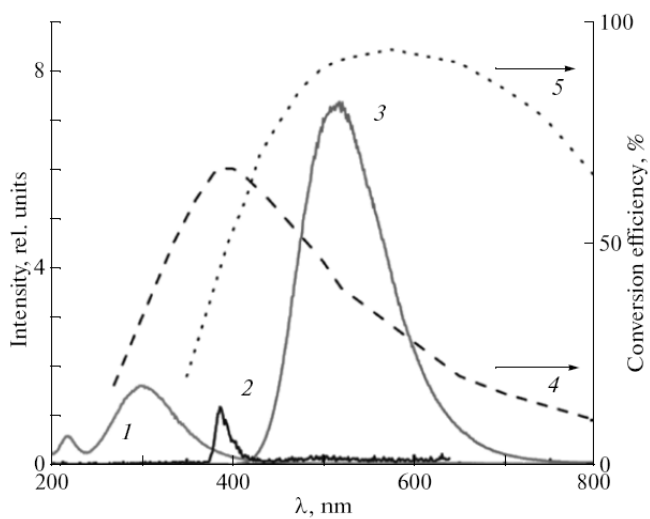

**Fig. 6.** X-ray luminescence spectra of (*1*) BaF2 single crystal, (*2*) ZnO:Ga ceramics, and (*3*) undoped ZnO ceramics at room temperature. Curves *4* and *5* correspond to typical conversion efficiencies of a SiPM and a CCD, respectively [60].

The integrated light yield of ZnO:Ga ceramics is ~5% of that of ZnO and ~25% of BaF2 (Fig. 6). The main reason for this low light yield is the low transparency of the scintillation ceramics in the range of the X-ray luminescence maximum at 386 nm. In [61], it was shown that, upon annealing in hydrogen atmosphere, the light yield of ZnO:Ga ceramics can be slightly improved. With allowance for the duration of the excitation pulse, the decay constant of the scintillation pulse of the ZnO:Ga ceramics is found to be 0.7 ns (Fig. 7b). According to the data of [59, 60], the relative light yield of ZnO ceramics in the energy range 35–662 keV varies within the limits of 2%, which is a good characteristic compared to other materials, e.g., NaI:Tl and LaBr3:Ce [62, 63].

Using the hydrostatic pressure technique, the authors of [8, 64, 65] succeeded in preparing ceramics based on pure ZnO and ZnO with addition of Al. As initial powders, they used materials synthesized under different conditions and having a significant scatter in the average size of grains. As a result, ceramics with a grain size in the range from 100 nm to 5 μm were obtained. In [64, 65], it was shown that the grain size and synthesis conditions substantially affect the optical and scintillation properties of ceramics. Although the decay times of the scintillation pulse were obtained as short (less than 2 ns), the authors of these works failed to reach a light yield comparable with that for monocrystalline ZnO. It is likely that the low light yield of ceramics is related to the low transparency of ceramic samples.

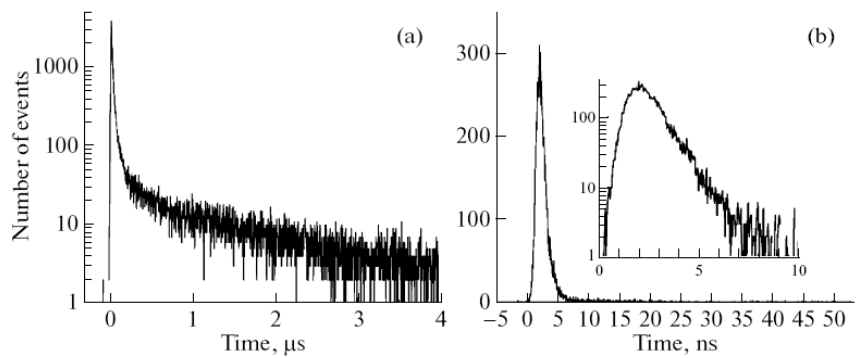

Fig. 7. Luminescence kinetics of (a) undoped and (b) gallium-doped ZnO ceramics at room temperature; the inset in (b) shows the kinetic curve during the first 10 ns (semilologarithmic scale). The decay constant τ (ns) is (a) 13 and $1.6 \times 10^3$ and (b) 1.1 [60].

Another group of authors studied the effect of boric acid on the synthesis parameters and luminescent properties of ZnO ceramics [66, 67]. They showed that, upon addition of up to 15% of boric acid to the initial powderlike ZnO, it becomes possible to achieve the effect of agglomeration of ceramic grains, to increase the conductivity, and to reduce the concentration of defects. These properties are the most important for production of varistors based on ZnO ceramics and, to a lesser extent, for improving the optical and scintillation characteristics. In [68], it was studied how the synthesis temperature affects the electric and luminescent properties of ZnO ceramics with Bi and Er additions, which are also used in varistors.

**CONCLUSIONS**

The fundamental absorption edge of zinc oxide can be shifted toward the short-wavelength spectral range by introducing Ga, In, and Mg impurities. As a rule, in these crystals, the intensity of the short-wavelength luminescence is higher than in pure ZnO. The edge luminescence associated with exciton and exciton-like states is observed in the near-UV spectral range and has subnanosecond deexcitation times. Embedding iron impurity makes it possible to reduce the edge luminescence decay time to tens of picoseconds. The position of the maximum of the edge luminescence can vary (within the limits of ~0.3 eV) depending on the form of the sample (powders, films, crystals), on the type of the impurity, and on the concentration of free carriers.

The luminescent properties of ZnO significantly depend on the perfection of the crystal structure and content of various defects in the crystal. Differences between experimental data and mechanisms of the ZnO luminescence are frequently determined by the fact that, in essence, researchers are investigating different objects. The properties of ZnO are also determined by the conditions and temperature of annealing of samples. For example, upon annealing of ZnO:N films at $T$ = 700–800°C, conductivity of the $p$-type is obtained, whereas annealing at $T$ > 900°C results in the conductivity of the $n$-type [23]. In crystals with anexcess of zinc, oxygen vacancies VO predominate, which serve as luminescence centers of the long-wavelength band. In crystals that were obtained in oxygen atmosphere, zinc vacancies VZn are luminescence centers. Discrepancies in the interpretation of the green luminescence are caused by the close positions of the maxima (~2.35 and ~2.5 eV for VZn- and VO-luminescence, respectively), as well as by a large width of these bands, $\Delta E_{1/2}$ > 300 meV. ZnO:Ga films, in which the intensity of the subnanosecond component is high, have already found application in scintillation detectors of ionizing radiations. Optical ceramics, the development of which is at the initial stage, has good prospects in this field.